# On the Design of a User-in-the-Loop Channel

## With Application to Emergency Egress

Constantin Dumitrescu

*Abstract*—The herein approach addresses the case when the user is part of the communication channel. A purpose of the abstraction is to provide a framework for the field of communications, with the user included in the system, reckoning the current 7 Layer model allows safeguards for data only. Solid grounds for application were identified as contribution to ensuring the need for safety which becomes tight during emergency events where timely evacuation is critical.

One of the components of a communication system [1] as described in information theory, is the communication channel. It allows the transmitter signal representing the message, to reach the receiver. A hardware setup with two different communication systems in a user-in-the-loop (UIL) [2] configuration is described and assessed pertinent to the specific application.

*Keywords—communication channel; user-in-the-loop; physical layer; wireless; layer 8; algedonic trigger; Internet of Things*

## I. INTRODUCTION

With more users striving to satisfy the higher level needs, and an increasing number of embedded applications easing the effort, a concatenated state of being safe requires continual balance to equilibrium, or mitigation either of acute and severe or prolonged and periodic scarcity of such stability.

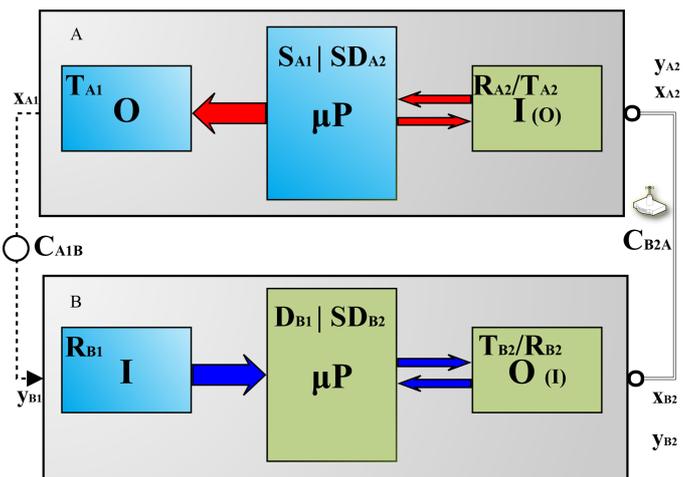

Fig. 1. Communication systems elements featuring UIL configuration shown in two different terminals.

In the former case, with the context of an infrastructure emergency egress of individuals from different forms of infrastructure assets (e.g. public buildings), the meeting of specified procedural requirements (e.g. timely evacuation) that require effective personal action is critical to safety and also leads to the fulfillment of target objectives for Safety Performance Indicators. The traditional method through which infrastructure administrators incentivize a peristatic [3] victory is the placement of visual indicators and indoor sketches on walls. Traditions are not being observed anymore, by most people.

The channel of a communication system relies on the first one of the 7 layers OSI (Open Systems Interconnect) model of computer networking, which is called the physical layer. The user's typical location is on top of the OSI stack, in Layer 8. When the channel model involves user interaction by means of a mobile communication system, it's storage device can be considered as a component of the loop through which a message is vehiculated towards a destination.

Such an attempt, with participating infrastructure-to-human communication, provides both parties with the traverse supporting awareness and goal reaching.

Insights from Information Theory research projects, at pace with the development of the Internet of Things (IoT) concept and applications, allow an ever boundless population of devices where such a participation could fit the purpose.

The setup in Fig. 1 shows the basic diagrams of two communication systems, in different colors and having the components prescribed by theory distributed among two devices/ terminals.

One of the terminals (device A) is a reference object intended to initiate communication and have no mobility. The second one (device B) is a mobile apparatus under direct control of the user.

A brief overview of the components and the intended application of the configuration, in theoretical terms, follows.

Let us consider the prime component of the communication loop as being the center component in device A, designated by µP (microprocessor unit) and colored in blue. It is the information source (designated by $S_{A1}$) in the first communication system ($Q_1$). The next component is the transmitter ($T_{A1}$), which is a device connected to an output port of the processing unit in device A.

Now the first channel ($C_{A1B}$) follows; it is enabled by direct user interaction with device B, which labels it as a user-in-the-loop channel. With regard to the application, it is also a one-way channel. This allows insight regarding the nature of $T_{A1}$ and some of it's characteristics.

For a full coverage of the initial assumptions, the expected features of the transmitter are listed below:

- no requirements to support high data rates;
- allows serial connection of multiple users;
- a high level of availability;
- actively promotes user awareness;
- low power consumption (able for battery operation);
- embeddable in a context.

Thus the elective space includes broadly two candidates for the interaction support medium: acoustical and optical.

With the first option it can be a more archaic sound producing device (e.g. a buzzer), but that reduces the applicability range since audible sound is also a source of interference in the user's Layer 8 noisy realm.

Hence, coin tossing with a single viable option, leads to choosing a simple LCD display for that purpose.

With the transmitter being of optical nature then the channel medium requirements arise: it should be free of obstacles that would prevent straight line propagation of light.

The receiver ($R_{B1}$) of the first communication system is located in device B. It's type shall be appropriate to the nature of the medium. It provides data to the input port of a processing unit ($D_{B1}$) found in device B and marked with µP. This unit is also the information source ($SD_{B2}$) in the second communication system ($Q_2$), and a destination of a message in the first system ($D_{B1}$). Data is sent to the output device (the second transmitter, $T_{B2}$) via an output port. Then the second channel follows ($C_{B2A}$); it does not have a UIL (user-in-the-loop) condition present, consequently no special requirements are specified – can be a one-way or two-way channel [4]. The next piece of the setup is the input component in device A, which is the receiver ($R_{A2}$) of the second communication system. It sends the data to the input port of the µP, making this also a destination ($SD_{A2}$) in the second system.

The $C_{A1B}$ channel medium matches the OSI physical layer of device A output and device B input, and the $C_{B2A}$ medium matches the physical layer at device B output and device A input. The first channel is rendered active/ inactive by user interaction (by controlling the input to device B); the second channel is always active, if enabled.

## II. A HARDWARE IMPLEMENTATION

The user interactivity is expressed by enabling/ disabling the first channel. This is most easily achieved if device B is a mobile device (e.g. smartphone/ tablet) with typical connectivity and imaging devices.

A camera with typical resolution and a device specific application able to analyze the pattern image and react appropriately to the decoded message ensure the receiving function over this type of channel. An internal nonvolatile memory buffer at device A provides storage of the data for subsequent processing and transmission.

The impetus promoting the user to scan the graphical pattern also drives the potential moving the data from $T_{A1}$ to $R_{B1}$ at the level of the physical layer.

A sample quick response pattern is shown in Fig. 2, with message encoded in version 3. This version allows 70 symbols to be included in the content of the message, enough to include network access, peer level signaling and other data.

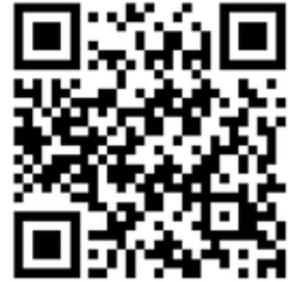

Fig. 2.   Quick Response pattern (version 3)

Device B is a smartphone with a camera and software application able to decode optically encoded information (i.e. quick response pattern).

Device A was designed as an info-display centered around a monochrome LCD with data fed by an 8-bit MCU (embedded USART and enough pins to deploy parallel data to the 128 by 128 dots graphical display). The high speed channel is centered on a WiFi solution, that also includes Ad-Hoc mode capabilities and UART communications port.

The info-display allows a Quick Response pattern version 3 [5] to be displayed (a 29 rows by 29 columns matrix and a quiet zone). Reconfiguring the display matrix allows lower versions of the standardized graphical pattern to be shown on the LCD.

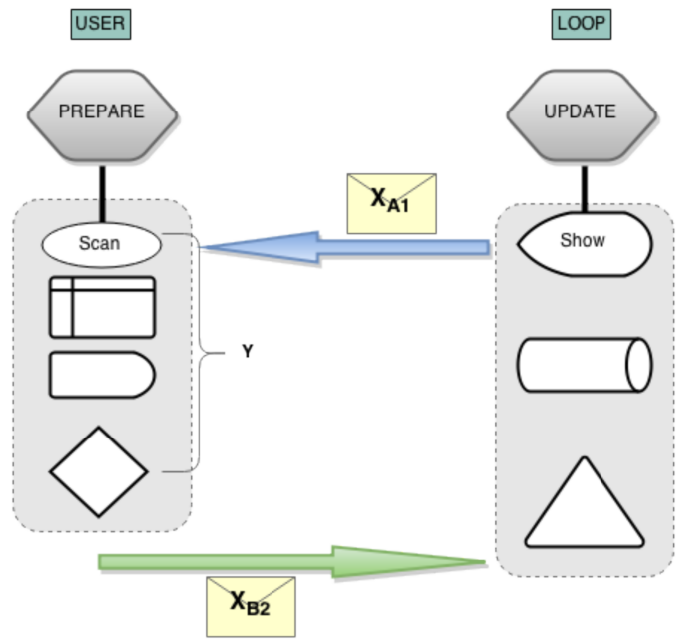

Fig. 3.   The simplified user-loop transaction.

The diagram in Fig. 3 depicts the decision point preceded by delay on the user mediated channel. Contrary to a usual channel, in the special case of a human user, this one involves a time delay larger than most data interface will tolerate, and usually incompatible with safety critical applications. The usage time for this type of channel should be minimal. Also a message obsolescence issue could be raised at the bilateral protocol message design stage.

The safe threshold of the LCD refresh rate with version 3 QR pattern, and a microcontroller running at 8 MHz clock speed with four cycles per instruction execution, was determined as 3 Hz. The low power design allows battery powered device operation, in continuous or safe mode.

## III. APPLICATION SPECIFIC ASPECTS

The onset of catastrophic event raises all risk levels to their top value. Auditory and visual annoyance alarms start at once. If the destination of the alert is the human being, the decision making unit within initiates the trivial sequence: "what's going on?"... By the time the individual figures out the urgency of the situation, the recommended egress period has already elapsed. An application of the UIL type of system described previously would imply the connection of the user to a communication network. When the infrastructure item (e.g. public building) is endowed with a viable system able to generate and update an algedonic [6] state vector, the users inside should be aware of it's availability and usability. The symbiotic flow of messages through the network involves a symbolic protocol to accommodate "Layer 8" expectancies (alert and guidance) with Layer 1 requirements (ensuring at least the initial handshake). The proposed method is to design the connection of the user to the high speed data channel using the message on the UIL channel itself, and to divide that operation in two steps:

- first encode a message with network access information, prior algedonic state vector, and close a transaction;

- then encode a message with the updated algedonic state vector, register the user on the network, upload specific data (indoor floor maps for evacuation purposes, instructions etc.), and close the second transaction.

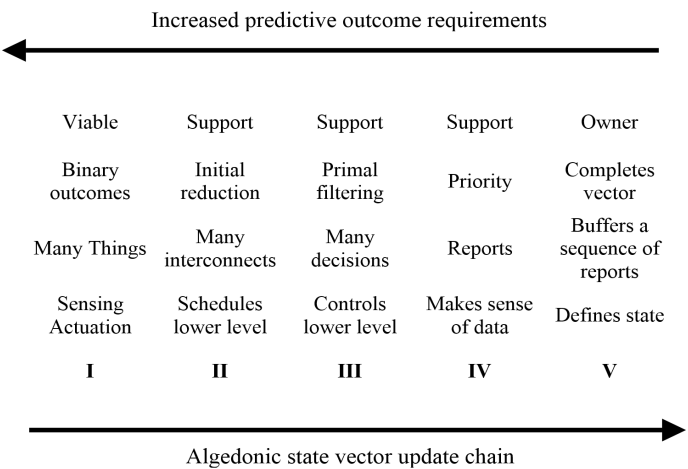

Fig. 4. Levels of algedonic state vector buildup.

With numerous architectural innovations the number of indoor or outdoor sensing and actuating functions is ever increasing. Thus, a large fan-in to an algedonic state vector update chain, requires a viable model system design, able to add meaning to the collected and processed data, and being aware that a non-random context involves a changing environment.

The algedonic state vector model is similar to the Viable System Model of organization. All lower level subsystems are support systems, except the ones at level I which generally are Internet of Things devices.

### A. Algedonic state vectors

The algedonic state is defined as an $n$ bit vector. The polling or sampling over specific time durations of various data flows in a network, the supporting and viable subsystems provide the information needed to update the vector, after being processed as required.

A 4 bit vector implies a minimum of four physical subsystems and at least four bits of buffer memory for the support systems.

### B. Alert trigger

When a "sufficiently pervasive device" like the human in the loop [7] is relying on the digital communications network to provide the support functions, the initial trigger to start the alert sequence is of most importance. In the evacuation egress scenario, it's statics and dynamics are very much different than the semaphore status of nowcasting weather events. Threshold determination seems like determining the lower "bound of silence" in a noisy environment. The trigger threshold itself, whether constant or adjustable, is raised when all algedonic state components have the same value. The type of alert initiated in either of the cases is outside the scope of the discussion.

A thorough analysis regarding the human in the loop vulnerability was performed by Lorrie Faith Cranor in [7], where impediments like environment variables render the communication inactive, eventually.

## IV. CONCLUSIONS

The number of IoT devices is growing constantly, with specific communication requirements between machines (M2M), machine-to-human (M2H), vehicle-to-infrastructure (V2I) being the focus of numerous research projects.

An introductory analysis of an infrastructure-to-human (I2H) type of communication with application to an emergency egress situation was performed in the context of the user-in-the-loop condition present in the one-way communication channel. This channel was modeled with message coding in the form of a graphical dot pattern displayed on an LCD screen. After the user scans the pattern using a mobile device a connection to a network is enabled through another channel, which also provides the return path, completing the loop.

Running simulated algedonic vectors in the looped configuration allows drawing some conclusions regarding the intended application of the design:

• a single pattern displaying device for each entry point of the infrastructure entity is easier to manage;

• the total number of users in the loop need not be known, provided the location and navigational signals can be provided;

• a 2.5 minutes target egress time is achievable with coordinated network management.

Further development of the application concept may include increased one-way connectivity at the displaying device (acoustic, light signaling/ LED), direct sensor reading and algorithms for message editing. The 8 bit microcontroller [8] allows both digital and analog sensor connections.

The WiFi module [9] power running modes may be controlled in order to save even more power, when extended coverage range is not required.

REFERENCES

[1] Claude E. Shannon. A mathematical theory of communication. Bell System Technical Journal, 27:379–423, 623–656, 1948.

[2] R. Schoenen, H. Yanikomeroglu, User-in-the-loop: spatial and temporal demand shaping for sustainable wireless networks, IEEE Communications Magazine, Volume 52, Issue 2, pp. 196-203, 2014.

[3] Danko Nikolić, Practopoiesis: Or how life fosters a mind, arXiv:1402.5332, 2014.

[4] Claude E. Shannon, Two-way Communication Channels. Proceedings of the Fourth Berkeley Symposium on Mathematical Statistics and Probability, Volume 1: Contributions to the Theory of Statistics, University of California Press, Berkeley, Calif., 1961, pp. 611-644.

[5] ISO/IEC 18004:2006, Information technology–Automatic identification and data capture techniques–QR Code 2005 bar code symbology specification, International Organization for Standardization, Geneva, Switzerland.

[6] Stafford Beer, The Heart of Enterprise, John Wiley, 1979.

[7] Lorrie Faith Cranor, A framework for reasoning about the human in the loop. In UPSEC'08: Proceedings of the 1st Conference on Usability, Psychology, and Security, pages 1–15, Berkeley, CA, USA, 2008. USENIX Association.

[8] DS39632E, PIC18F2455/ 2550/ 4455/ 4550 Data Sheet, Microchip Technology Inc., 2009.

[9] DS70005171A, RN171, 2.4GHz IEEE Std. 802.11 b/g Wireless LAN Module.